\documentclass[aps,prl,twocolumn,superscriptaddress,showpacs]{revtex4}
\usepackage{amsmath}
\usepackage{graphicx}
\begin{document}
\title{High temperature thermal conductivity of 2-leg spin-1/2 ladders}
\author{X. Zotos}

\affiliation{Institut Romand de Recherche Num\'erique en Physique des
Mat\'eriaux (IRRMA),\\PPH-Ecublens, CH-1015 Lausanne, Switzerland }
\date{\today}
\begin{abstract}
Based on numerical simulations, a study of the high temperature,   
finite frequency, thermal conductivity $\kappa(\omega)$ 
of spin-1/2 ladders is presented. 
The exact diagonalization and a novel Lanczos technique 
are employed.The conductivity spectra,  
analyzed as a function of rung coupling, point to a non-diverging 
$dc-$limit but to an unconventional low frequency behavior.
The results are discussed with perspective recent experiments indicating 
a significant magnetic contribution to the energy transport in quasi-one 
dimensional compounds.
\end{abstract}
\pacs{66.70.+f, 75.10.Pq, 75.40.Gb, ,75.40.Mg} 

\maketitle
{\it Introduction ---}
Recent experiments \cite{ott1,ott2,buch,kudo,ott3} convincingly 
promoted magnetic excitations as a very efficient mechanism for energy 
transport in quasi-one dimensional materials.  
In particular, spin-1/2 Heisenberg chain and ladder compounds  
(undoped or hole doped) have been studied. 
In both systems, the observed highly anisotropic 
thermal conductivity (comparable in magnitude to that of 
metallic systems) was attributed to magnetic transport.

Theoretically, it has been noticed that  
in the one dimensional (1D) spin-1/2 Heisenberg model the energy current 
commutes with the Hamiltonian and thus the thermal conductivity is 
ballistic at all temperatures \cite{huber,nie,znp}. This observation 
falls in line with a proposal of unconventional 
transport in 1D integrable systems \cite{znp}.
Besides this rigorous result, the experimental developments  
motivated theoretical studies of the thermal conductivity in 
other (e.g. frustrated, gapped) spin chain and ladder Hamiltonians 
based on numerical simulations \cite{gros,brenig,brenig2} or 
low energy effective theories \cite{orignac,saito}. 
These works focused on the thermal Drude weight
$D_{th}$ as a criterion of ballistic transport \cite{znp,naef,klumper} 
and they lead to an active discussion whether nonintegrable 
systems also show ideal thermal conductivity.

The key to obtaining reliable information by numerical simulations 
is the study of large enough systems that allow analysis of the scaling 
with lattice size.
In this work, we study the thermal conductivity  within 
linear response theory, first,    
at high temperatures where numerical simulations on finite systems  
are most reliable (the characteristic thermal scattering length 
should be smaller than the size of the lattice) and second,
by using a novel technique based on the Lanczos method and the microcanonical 
ensemble (MCLM \cite{mclm}). This method gives access to  
fairly larger systems than those studied till now by the
exact diagonalization (ED) method.
 
The main issues of actual experimental and theoretical interest that 
we address are: 
(i) whether the thermal transport is ballistic or diffusive, 
(ii) how the interchain coupling affects the ideal thermal transport 
of the decoupled (integrable) Heisenberg spin-1/2 chains, 
(iii) what is the order of magnitude of the magnetic contribution to the 
thermal conductivity as a function of exchange couplings and in particular 
how it compares with experimental values in ladder compounds.

\bigskip
{\it Hamiltonian and Method ---}
The 2-leg ladder Hamiltonian is given by the $q=0$ component
of $H_q$, the Fourier transform of the local energy density, that we 
define as,
\begin{eqnarray}
H_q&=&J\sum_{l=1,L} e^{iq(l+1/2)} ({\bf S}_{1,l+1} \cdot {\bf S}_{1,l} +
{\bf S}_{2,l+1} \cdot {\bf S}_{2,l})
\nonumber\\
&+&\frac{J_{\perp}}{J} e^{iql} {\bf S}_{1,l} \cdot {\bf S}_{2,l}.
\label{hq}
\end{eqnarray}
\noindent
${\bf S}_l$ are spin-1/2 operators at site $l$ and in the following, 
we consider systems with periodic boundary conditions, 
we take $J=1$ as the unit of energy 
($\hbar=k_B=1$) and unit lattice constants.

To study the thermal conductivity we define the energy current 
operator $j^E$ using the continuity equation for $H_q$ \cite{hardy,nonlocal}. 
Extracting the 
long-wavelength transport limit,
$\partial H_q/\partial t \sim -iq j^E$  for $q\rightarrow 0$, we obtain, 
\begin{eqnarray}
j^E&=&J^2 i \sum_{l=1,L} \{
{\bf S}_{1,l-1}\cdot ({\bf S}_{1,l} \times {\bf S}_{1,l+1})
\\
&+&{\bf S}_{2,l-1}\cdot ({\bf S}_{2,l} \times {\bf S}_{2,l+1})\}
\nonumber\\
&+& \frac{J_{\perp}}{2J} \{
  {\bf S}_{1,l-1}\cdot ({\bf S}_{1,l} \times {\bf S}_{2,l})
+ {\bf S}_{2,l-1}\cdot ({\bf S}_{2,l} \times {\bf S}_{1,l})
\nonumber\\
&+& {\bf S}_{2,l}\cdot ({\bf S}_{1,l} \times {\bf S}_{1,l+1})
+ {\bf S}_{1,l}\cdot ({\bf S}_{2,l} \times {\bf S}_{2,l+1}) \}. 
\nonumber
\label{je}
\end{eqnarray}

Within linear response theory \cite{lutt}, 
the real part of the thermal conductivity at frequency $\omega$ is given by,
\begin{equation}
\kappa(\omega)=2\pi D_{th}\delta(\omega) +\kappa_{reg}(\omega),
\label{kappa}
\end{equation}

\noindent
with the regular part,
\begin{equation}
\kappa_{reg}(\omega >0)=\frac{\beta}{\omega L}\tanh(\frac{\beta\omega}{2}) 
\Im i \int_{0}^{+\infty} dt e^{izt} \langle \{ j^E(t),j^E \} \rangle,
\label{kappar}
\end{equation}

\noindent
$(\beta=1/T,~z=\omega+i\eta)$ and the thermal Drude weight,  
\begin{equation}
D_{th}=\frac{\beta^2}{2L}\sum_n p_n  
\mid \langle n\mid j^E \mid n \rangle \mid^2.
\label{drude}
\end{equation}

\noindent
$\mid n\rangle$ are the eigenstates and $p_n$ the Boltzmann weights.
For $\beta\rightarrow 0$ we can deduce the sum-rule,
\begin{equation} 
\int_{-\infty}^{+\infty} d\omega \kappa(\omega)=\frac{\pi \beta^2}{L}
\langle {j^E}^2 \rangle=I,
\label{sum}
\end{equation}

\noindent
that suggests the analysis of $2\pi D_{th}/I$ in order to estimate the 
contribution of ballistic transport.

In the following, we employ the ED method to study 
$D_{th}$ and $\kappa({\omega})$ for systems up to $L=9$ rungs. 
For larger systems, up to $L=14$, we use the MCLM method. In both 
cases we implement the translational symmetry that provides 
independent $k-$Hamiltonian subspaces. In the MCLM  
method \cite{mclm}, we replace the thermal average in  
(\ref{kappar}) by the expectation value over a single state 
$\mid \lambda \rangle$ of energy 
$\lambda$ that equals the canonical ensemble value of the energy at the 
desired temperature \cite{ll}, $\lambda=\langle H \rangle$. 
In most cases, we present results for $\beta \rightarrow 0$  
where $\lambda \sim 0$ due to the symmetric spectrum of the Hamiltonian. 
Otherwise, we determine $\lambda$ by extrapolation of thermal energies 
evaluated using ED results.

The state $\mid \lambda \rangle$ is constructed by employing a 1st Lanczos 
procedure of about $1000$ steps using as ``effective Hamiltonian'' the 
operator $K=(H-\lambda)^2$; in practice, for large systems with dense 
spectra, the procedure cannot fully converge. $\mid\lambda\rangle$,  
the ground state of $K$ in the constructed Lanczos subspace, 
is  characterized by a distribution over the eigenstates $\mid n\rangle$ 
of variance in energy of $O(0.01)$ that imposes a maximum $\omega$ 
resolution to the spectra. Then, starting from $j^E \mid\lambda\rangle$, 
a 2nd Lanczos procedure of about $4000$ steps provides  
$\kappa(\omega)$ using the continued fraction technique \cite{cfe} 
(typically $\eta \sim 0.01$). Thus we effectively 
evaluate $\kappa(\omega)$ by,
\begin{eqnarray}
\kappa(\omega)&\mapsto& \frac{\beta}{\omega L}\tanh(\frac{\beta\omega}{2})
(-\Im) (\langle \lambda\mid j^E\frac{1}{z-(H-\lambda)}j^E\mid\lambda\rangle
\nonumber\\
&+& \langle \lambda\mid j^E\frac{1}{z+(H-\lambda)}j^E\mid\lambda\rangle).
\end{eqnarray}

\noindent
Note that, as the Lanczos procedures do not fully converge, an eventual 
$\delta(\omega)$-peak contribution appears as a broadened weight 
at very low frequencies \cite{mclm}.

\begin{figure}
\includegraphics[width=8.0cm]{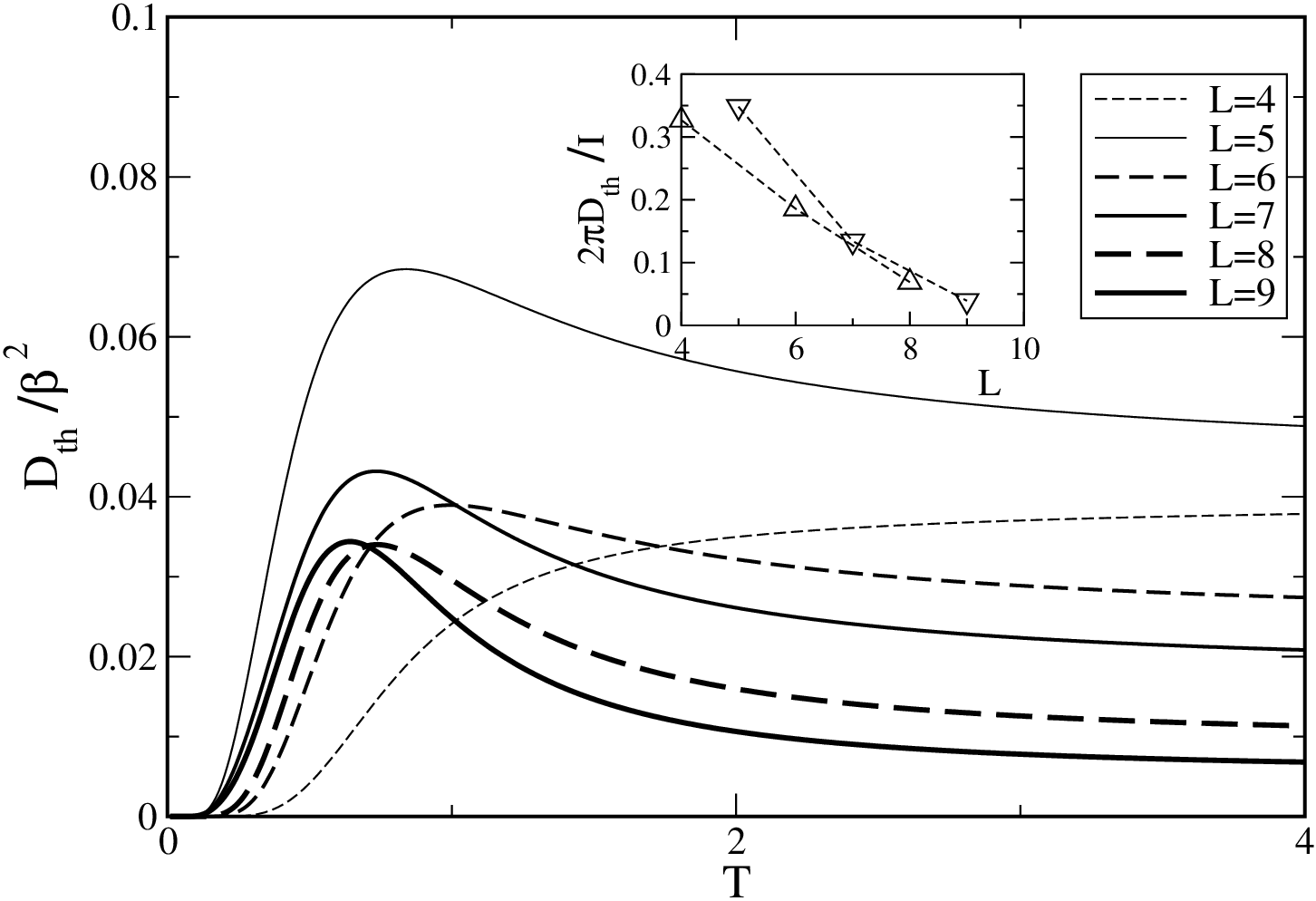}
\caption{Thermal Drude weight for $J_{\perp}=1$ as a function 
of temperature for systems with $L$ rungs (ED evaluation). 
Inset: normalized $D_{th}$ for $\beta\rightarrow 0$ as a function of system 
size.}
\label{fig1}
\end{figure}

\bigskip
{\it Thermal conductivity ---}
The first step in characterizing $\kappa(\omega)$ is the 
evaluation of the Drude weight at finite $T$. If $D_{th}$ 
is finite then the transport is ballistic; if it vanishes, 
the transport is normal provided the    
$\kappa_{dc}=\kappa(\omega \rightarrow 0)$ limit exists.
For a finite system $D_{th}$ is always nonzero, so it is crucial  
to examine its scaling as a function of size. Exact results obtained 
by the ED method are shown in Figure \ref{fig1}. In the high 
temperature limit $D_{th}/\beta^2$ rapidly decreases, seemingly exponentially 
fast, with system size (for each series of even or odd $L$). 
It already represents only a couple of percent of 
the sum-rule for $L=9$ rungs (inset). 
Of course, the size of the studied lattices is rather limited but a vanishing 
Drude weight in the $\beta\rightarrow 0$ limit is corroborated 
by the $\kappa(\omega)$ spectra shown below.  

At lower temperatures we find a non-monotonic scaling  
for even-rung systems. Note however that the 
change of scaling behavior of $D_{th}$,  
from decreasing with increasing $L$ at high-$T$ 
to the opposite at low-$T$, shifts to lower temperature as we 
consider larger systems (see also \cite{brenig2}).
We can then argue that over the whole temperature range the Drude weight 
scales to zero as $L\rightarrow \infty$.
It would be unexpected if there is a transition to a finite $D_{th}$ 
below some critical temperature. 

Next, we discuss the thermal conductivity as 
a function of $\omega$ and $L$
in an attempt to determine whether it represents normal 
transport (diffusive behavior exemplified by a Lorentzian form),  
unconventional (e.g. power law at low frequencies) or some other behavior 
that still implies a finite $dc-$conductivity. Furthermore, we study the 
dependence on $J_{\perp}$ in order to 
find out the effect of interchain coupling on the ballistic transport 
of decoupled chains. For $J_{\perp}=0$ only the $2\pi \delta(\omega) D_{th}$ 
contribution exists as $[j^E,H]=0$.
\begin{figure}
\includegraphics[width=8.0cm]{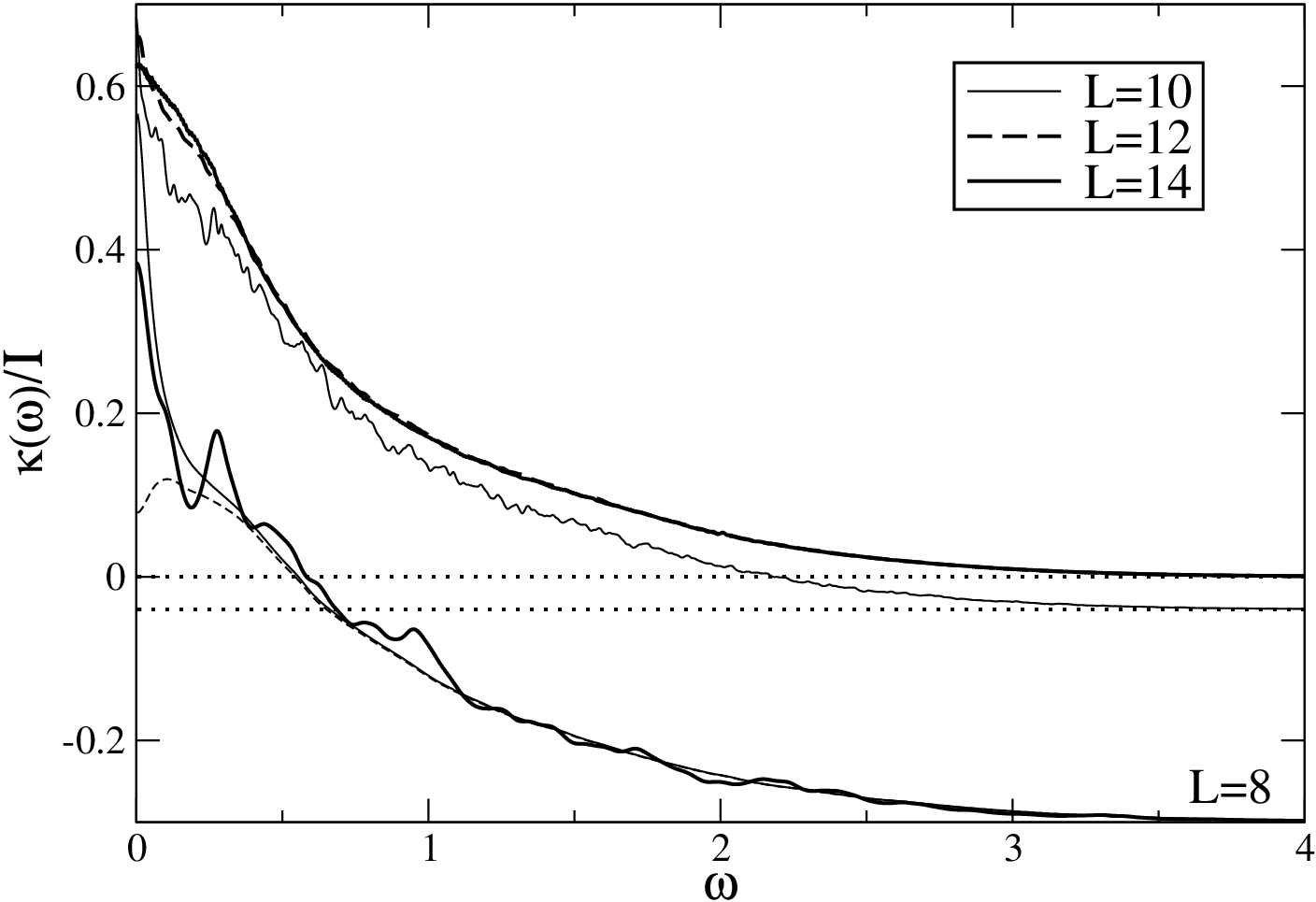}
\caption{Normalized thermal conductivity for different $L$'s by the 
MCLM method (curves displaced for clarity). 
ED evaluation for $L=8$: thin line - $\kappa(\omega)/I$, 
thin dashed line - $\kappa_{reg}(\omega)/I$.}
\label{fig2}
\end{figure} 

In Figure \ref{fig2} we show a series of spectra for different $L$'s, 
$J_{\perp}=1$ and $\beta\rightarrow 0$. 
To illustrate the applicability of the MCLM method,   
we compare a spectrum from a canonical ensemble ED study 
for $L=8$ (weights smoothed with Lorentzians) with one 
obtained by the MCLM averaged over all $k-$symmetry subspaces 
(in both cases $\eta=0.05$). We find that the agreement is fair 
and the broadened Drude $\delta(\omega)-$peak is reproduced.  
We also show the regular part from the ED evaluation indicating that 
the presence of the Drude weight, that is fairly large for $L=8$, 
is limited to $\omega\le 0.3$.

Next, in order to examine the finite size scaling of 
$\kappa(\omega)$, we present 
spectra for larger systems, $L=10,12,14$ ($\eta=0.01$). 
We notice that the statistical fluctuations  
drastically decrease with increasing $L$ so that for $L=14$ it is 
sufficient to consider only one $k-$subspace ($k=0$ is shown, the curves 
being practically indistinguishable for the other $k-$subspaces). 
The reason is that the dimension of the Hilbert 
space increases by a factor $\sim 10$ when increasing $L$ 
by 2 rungs. For $L=14$ it is $O(3\times 10^6)$ states in each 
$k-$subspace and so it represents a fairly dense spectrum at high 
energies (temperatures). Thus, in the following, we show spectra  
constructed with only the $k=0$ subspace for $L=14$ rungs.

Finally, we note that the spectra rapidly collapse to the same curve. 
Noticeable deviations 
for $L \ge 10$ are limited to the $\omega\sim 0 - 0.3$ range. 
The data apparently converge to a $\kappa(\omega)$ curve that:  
(i) it is characterized by a finite $\kappa_{dc}=\kappa(\omega\rightarrow 0)$ 
value, 
(ii) it approaches the $\omega\rightarrow 0$ limit with a finite slope, 
(iii) it shows a change of curvature at $\omega\sim 0.5$ for all $L$, 
(iv) its overall $\omega$ dependence does not correspond to a 
Lorentzian or power law behavior; a minimal description of $\kappa(\omega)$ 
is obtained by an $e^{-\mid \omega\mid\tau}$ form \cite{multi}.

Of course, the deduced $\omega\rightarrow 0$ behavior is  
rather tentative considering the limited size of the systems that we are able 
to study. A finite $L$ imposes a cutoff on the lowest 
frequency behavior that can be reliably extracted. We cannot 
exclude the emergence of a more conventional form  - with vanishing 
zero frequency slope - in the range $\omega \sim 0 - 0.2$.
The apparent finite slope might also be due to the mixing, in the 
very low frequency spectrum, of a remnant $\delta(\omega)$-Drude 
contribution that we expect to dissapear as $L\rightarrow \infty$ 
(see Figure \ref{fig1}).
In any case, a prominent Drude peak is not observed.
\begin{figure}
\includegraphics[width=8.0cm]{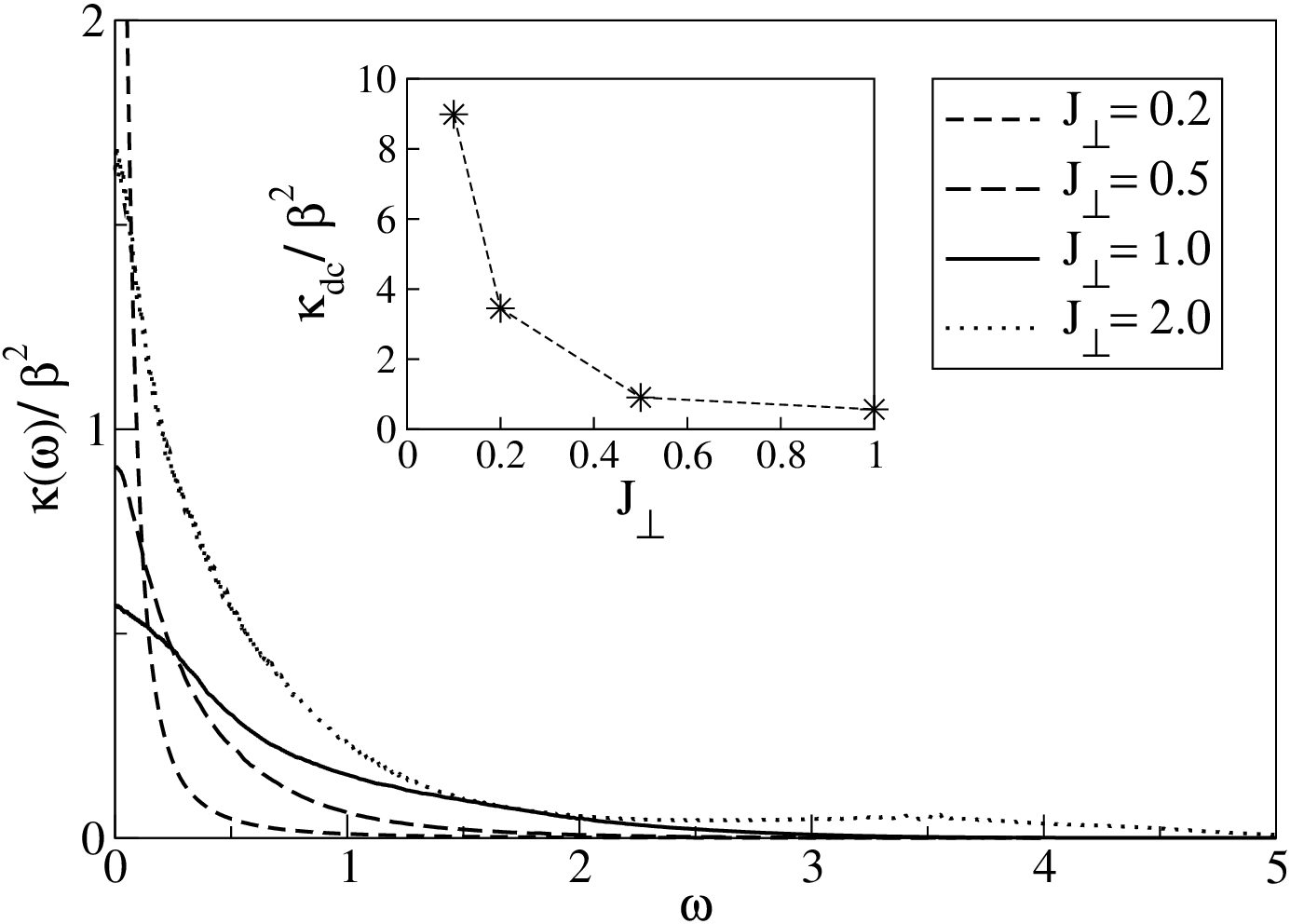}
\caption{(a) Thermal conductivity for $L=14$ rungs as a function of 
$J_{\perp}$ for $\beta\rightarrow 0$. Inset: $dc-$limit.} 
\label{fig3}
\end{figure} 

Next, in Figure \ref{fig3}, we show a series of spectra 
for $L=14$ as a function of interchain coupling $J_{\perp}$ 
in the $\beta\rightarrow 0$ limit. For $J_{\perp} < 1$, first, 
the $dc-$conductivity (inset) scales as $\kappa_{dc}\sim 1/J_{\perp}^2$,  
second, $\kappa(\omega)$ correctly tends to 
a $\delta(\omega)-$peak as $J_{\perp} \rightarrow 0$ signaling the 
ballistic transport of decoupled chains, third, 
the frequency range of finite weight extends up to 
$\omega \sim 4J_{\perp}$. For $J_{\perp}>1$, a second weak peak appears at  
$\omega \sim 2J_{\perp}$ (corresponding 
to the energy of singlet-triplet rung excitations) and 
$\kappa_{dc}\sim J_{\perp}$ as now $J_{\perp}$ becomes now the dominant 
energy scale.
Notice that for $J_{\perp} < 1$, the overall behavior 
of $\kappa(\omega)$ is what one would 
qualitatively expect by taking into account within a perturbative scheme 
the effect of interchain coupling on the ballistic transport of 
a non-interacting system. In this case however, the ``free" limit is 
the spin-1/2 Heisenberg Hamiltonian which implies that 
the effect of a perturbation on the ideal transport of an integrable 
system is described within a simple picture.
\begin{figure}
\includegraphics[width=8.0cm]{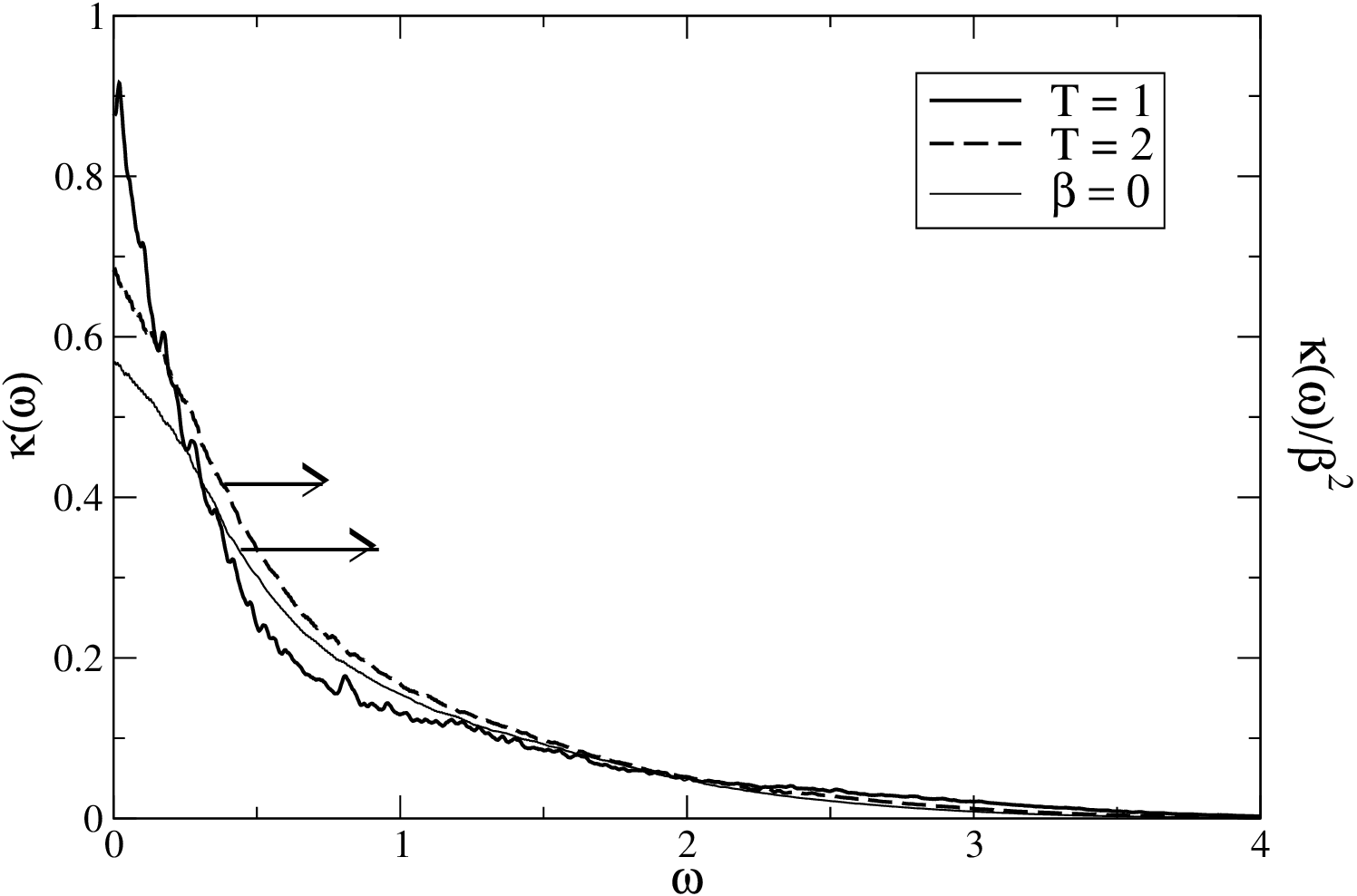}
\caption{Thermal conductivity for $L=14$ at $T=1$, $T=2 (\beta=0.5)$ 
and for comparison at $\beta\rightarrow 0$.}
\label{fig4}
\end{figure} 

In Figure \ref{fig4} we show the frequency dependence of the 
thermal conductivity at lower temperatures. 
An overall similar behavior as for $\beta\rightarrow 0$ is found except 
that the rounding at $\omega\sim 0.5$ disappears for $T=1$.
The data are qualitatively described by two separate exponential forms, 
one in the frequency range $\omega\sim 0 - 0.7$ and another one above.
No excessive enhancement of the $dc-$conductivity is observed. 
The sum-rule (\ref{sum}) is now satisfied to about 95 percent and 
the $L$ dependence of the conductivity spectra remains within the  
statistical noise of the data.
It is rather difficult to study even lower temperatures as the energy spectrum 
is very sparse at low energies and thus the fluctuations significantly 
increase. 

From the extracted dependence of $\kappa_{dc}$ on $J_{\perp}$ 
we can estimate an experimentally relevant - $\kappa^{exp}_{dc}$ - 
magnetic contribution to the thermal conductivity at high temperatures. 
Re-inserting units in the basic expression (\ref{kappar}), 
$\kappa^{exp}_{dc}$ is given by,

\begin{equation}
\kappa^{exp}_{dc}=
\frac{k_B}{abc} (\frac{J^2 c}{\hbar})^2 (\frac{\hbar}{J})\frac{1}{J^2}
\kappa_{dc}(\beta J,\frac{\hbar}{J}\omega).
\end{equation}
\noindent
$a,b,c$ are lattice constants ($c$ along the ladder axis) and 
energies are in degrees K.
For characteristic lattice constants of $O(10 \AA)$ and $J\sim O(1000$K), 
we obtain a thermal conductivity $O(2$ W/mK) that scales as 
$(\beta J)^2$ and $(J/J_{\perp})^2$ for $T\gg J $.
If we assume an exponential increase of the conductivity 
(due to the freezing of Umklapp processes) 
with characteristic energy $J$ below a temperature of $O(J)$, 
we obtain a room temperature thermal conductivity of order 10 - 100 W/mK 
that is consistent with experiment. 
We should keep in mind in this estimation that it is 
notoriously difficult \cite{berman} to establish the temperature dependence 
of the thermal conductivity for $T \le J$.

We can then argue, that the experimentally observed high values of the 
thermal conductivity \cite{ott1,buch} in ladder compounds 
are due to the large exchange coupling $J$ and that they are 
limited by spin-spin as well as spin-phonon scattering. 
This is in contrast to quasi-1D materials described 
by the Heisenberg spin-1/2 Hamiltonian \cite{ott2,ott3} where it is 
unambiguous that the spin-spin scattering is absent.

\bigskip
{\it Discussion -}
The presented numerical study indicates that, at least in the high 
temperature limit, the thermal conductivity 
in 2-leg spin-1/2 ladders is characterized by a finite 
$dc-$value and thus non ballistic transport.  
The low frequency spectra do not seem consistent with a 
Lorentzian form. At lower temperatures there is a change in behavior but 
not a dramatic enhancement of the $dc-$conductivity or the appearance of a 
prominent Drude peak. Presently, with the available numerical methods, 
it is rather challenging to explore an eventual crossover to a 
quasi-ballistic regime at low temperatures.

Furthermore, this analysis exemplifies the effect of a non-integrable 
interaction (interchain coupling) 
on the ballistic transport of an integrable system. 
The obtained perturbative result indicates that the presence in a Hamiltonian 
of an integrable case with diverging conductivity  
is signaled over a finite range in interaction parameter space. 

Regarding experimental realizations, this study is more relevant to 
ladder compounds with $J$ smaller than the room temperature but it shows   
that materials with high thermal conductivities can be obtained by increasing 
the coupling $J$ and/or reducing the interchain coupling $J_{\perp}$. 
It also shows that it would be interesting to explore the unusual frequency 
dependence, for instance, by light scattering.
As to the observability of the exceptional transport in integrable systems,  
we expect the effect to be most remarkable at high temperatures in 
small exchange coupling $J$ one dimensional spin-1/2 Heisenberg systems.

\bigskip
{\it Acknowledgments ---}
It is a pleasure to thank M.Long, P. Prelov\v sek, C. Gros, J. Karadamoglou, 
for useful discussions and acknowledge financial 
support by the Swiss National Science Foundation, the  
University of Fribourg and the University of Neuch\^atel.


\begin{references}
\bibitem{nonlocal} We do not use the continuity equation for the local 
energy as it might lead to ambiguities due to the nonlocality of the 
energy operator.
\end{references}
\end{document}